\pdfoutput=1  
\documentclass[11pt,a4paper]{article}

\usepackage[T1]{fontenc}
\usepackage[utf8]{inputenc}
\usepackage{amsmath,amssymb}
\usepackage{graphicx}
\usepackage{bm}
\usepackage{listings}
\usepackage{xcolor}
\usepackage{booktabs}
\usepackage{longtable}
\usepackage{hyperref}

\makeatletter
  \usepackage[margin=2.5cm]{geometry}%
  \newcommand{\@instituteinfo}{}%
  \newcommand{\institute}[1]{\gdef\@instituteinfo{#1}}%
  \newcommand{\@abstracttext}{}%
  \renewcommand{\abstract}[1]{\gdef\@abstracttext{#1}}%
  \newcommand{\PACS}[1]{}%
  \newcommand{\inst}[1]{\textsuperscript{#1}}%
  \newcommand{\journalname}[1]{}%
  \let\svj@oldmaketitle\maketitle
  \renewcommand{\maketitle}{%
    \svj@oldmaketitle
    \begingroup\let\and\par
      \begin{center}\small\itshape\@instituteinfo\end{center}%
    \endgroup
    \begin{quotation}\noindent\small\@abstracttext\end{quotation}%
    \bigskip}%
\makeatother

\journalname{Eur. Phys. J. A}

\DeclareRobustCommand{\ket}[1]{\left|#1\right\rangle}

\providecommand{\Id}{\hat{\mathbf 1}}
\providecommand{\braket}[2]{\left\langle #1 \middle| #2 \right\rangle}
\providecommand{\me}[3]{\left\langle #1 \middle| #2 \middle| #3 \right\rangle}
\DeclareRobustCommand{\iso}[5]{\left\langle #1\, #2 \middle| #3\, #4\right\rangle^{#5}}
\DeclareRobustCommand{\su}[1]{\mathrm{#1}}
\DeclareRobustCommand{\lm}{(\lambda,\mu)}
\DeclareRobustCommand{\code}[1]{\texttt{#1}}
\DeclareRobustCommand{\Su}{\code{Su3cgvcs}}

\begin{document}

\title{SU3HOB-ladder: a Fortran package for harmonic-oscillator brackets in
       the SU(3) basis from pseudo-spin ladder operators\thanks{Intended for the
       \emph{Tools for Experiment and Theory} section of Eur. Phys. J. A.}}

\author{A.~Step\v{s}ys\inst{1}\thanks{\emph{e-mail:} augustinas.stepsys@ftmc.lt}
        \and S.~Mickevi\v{c}ius\inst{2}
        \and D.~Germanas\inst{1}}

\institute{Center for Physical Sciences and Technology (FTMC),
           Saul\.etekio al.~3, LT-10257 Vilnius, Lithuania
           \and
           HEI Kauno kolegija, Pramon\.es ave.~20, 50468 Kaunas, Lithuania}

\date{}

\abstract{We present \code{SU3HOB-ladder}, a Fortran~2008 package computing the general
Talmi--Moshinsky harmonic-oscillator transformation brackets for an arbitrary
mass-ratio parameter $d$ in the $\su{SU}(3)$ basis. The bracket reduces to a
single sum over Wigner $d$-functions of the reflection-containing
Talmi--Moshinsky matrix, weighted by products of
$\su{SU}(3)\supset\su{SO}(3)$ isofactors of the chain
$\su{U}(6)\supset\su{U}(3)\times\su{U}(2)$; the isofactors are
$d$-independent, so they are built once per block $(E,L)$ and reused for every
$d$. We obtain them from the $\su{U}(2)$ pseudo-spin generators of that same
chain: because the $\su{U}(3)$ and $\su{U}(2)$ labels inside $[E]$ of
$\su{U}(6)$ are complementary, the highest-weight states are the null space of
the raising operator $\hat J_{+}=\mathbf a^{\dagger}(1)\cdot\mathbf a(2)$ and
the remaining members of each multiplet follow by lowering, so that no
$\su{SU}(3)$ recoupling machinery, $K$-matrix or outer-multiplicity resolution
is required at any point. We compare this construction with two alternatives
that produce the same brackets --- the same scheme with isofactors taken from
the vector-coherent-state library \Su{}~\cite{Bahri_2004}, and the classical
Talmi--Moshinsky sum of Kamuntavi\v{c}ius et al.~\cite{Kamuntavicius} --- over
every block up to $E=50$. Evaluating the classical sum with its factorials in
logarithmic form rather than as tabulated binomials removes the range limit
that otherwise caps it near $E\simeq12$, and makes it an absolute check on the
bracket values over the whole range where its cost is bearable. Against that
check the two routes agree to within a factor of a few at every shell, while
the ladder construction is faster by three orders of magnitude; it is also the
only one of the three usable over the whole range, retaining
$\max|\mathrm{HOB}\cdot\mathrm{HOB}^{T}-I|\lesssim2\times10^{-6}$ at $E=50$,
where the library route has become unusable and the classical route is
prohibitively slow.
}

\maketitle

\section*{Program summary}
\noindent
\begin{tabular}{@{}p{0.30\linewidth}p{0.64\linewidth}@{}}
\emph{Program title:} & \code{SU3HOB-ladder}\\[2pt]
\emph{Version:} & 1.0\\[2pt]
\emph{License:} & MIT\\[2pt]
\emph{Repository / DOI:} &
  \url{https://github.com/Augustinaz/SU3HOB-ladder}; archived at
  \href{https://doi.org/10.5281/zenodo.22150397}{10.5281/zenodo.22150397}
  (concept DOI, all versions)
  \\[2pt]
\emph{Programming language:} & Fortran~2008\\[2pt]
\emph{External dependencies:} & The SU(2) Clebsch--Gordan coefficients needed
  for the $\hat J_{+}$ matrix elements are taken from the \code{WignerSymbol}
  library~\cite{WignerSymbol} (MIT License), bundled unmodified as a single
  module so that the package builds standalone. The validation driver
  evaluates the classical Talmi--Moshinsky bracket of Kamuntavi\v{c}ius et
  al.~\cite{Kamuntavicius} as an independent reference
  (\code{tmb\_kam.f90}). BLAS/LAPACK (\code{dsyev}) is used only by the
  benchmark driver. An
  optional build path substitutes the external Clebsch--Gordan library
  \Su{}~\cite{Bahri_2004} for the isofactor step; it is not required, and is
  used here only for validation.\\[2pt]
\emph{Nature of problem:} & The general Talmi--Moshinsky harmonic-oscillator
  bracket relates two two-oscillator product states coupled to total orbital
  angular momentum $L$ under an orthogonal transformation of the Jacobi
  coordinates with arbitrary mass-ratio parameter $d$. Brackets are the
  elementary ingredient of translationally invariant few-body and no-core
  shell-model matrix elements, and are needed for many blocks $(E,L)$ and many
  values of $d$.\\[2pt]
\emph{Solution method:} & The bracket is reduced to Eq.~(\ref{eq:hob}): a sum
  over $J$ of Wigner $d$-functions of the reflection-containing
  Talmi--Moshinsky matrix, weighted by products of
  $\su{SU}(3)\supset\su{SO}(3)$ isofactors. The isofactors are obtained from
  the $\su{U}(2)$ pseudo-spin generators of the same group chain: the
  highest-weight states are the null space of $\hat J_{+}$ in one block, and
  the remaining members of each multiplet follow by lowering
  (Sec.~\ref{sec:raising}). No $\su{SU}(3)$ recoupling machinery, $K$-matrix
  or outer-multiplicity resolution is required. The towers are independent of
  $d$ and are built once per $(E,L)$.\\[2pt]
\emph{Restrictions:} & Verified over every block with $E\le50$ ($1301$ blocks,
  largest $2907\times2907$, $1.26\times10^{9}$ brackets in all). Accuracy is
  limited by the mutual orthogonality of different $J$ towers, which is exact
  by symmetry and degrades only through rounding; see
  Sec.~\ref{sec:cost} and Table~\ref{tab:accuracy}.\\[2pt]
\emph{Typical running time:} & All $1301$ blocks with $E\le50$: $78$~s for the
  brackets on one core (prepare plus evaluate). A single $E=50$ shell: $13$~s
  (Table~\ref{tab:speed}).\\[2pt]
\emph{Computing environment:} & All timings quoted in this paper are
  single-core on one x86-64 desktop CPU, \code{gfortran}~14.2 with
  \code{-O2}, OpenBLAS 0.3.29 pinned to one thread. The library itself calls
  no external linear algebra: the Gram matrices of Sec.~\ref{sec:raising} are
  diagonalized by its own cyclic Jacobi routine. \code{dsyev} appears only in
  the drivers, where it diagonalizes $\hat\Lambda$ for the coefficients of
  fractional parentage; the diagonalization column of Table~\ref{tab:cfp} is
  therefore an OpenBLAS \code{dsyev} timing, while every ladder timing in this
  paper is Jacobi.\\
\end{tabular}

\section*{Note on this version}
Version~1 of this arXiv entry was a different, shorter paper by the same
authors: \emph{\code{SU3HOB-cgvcs}: harmonic-oscillator brackets in the
$\su{SU}(3)$ basis with isofactors from an external
$\su{SU}(3)\supset\su{SO}(3)$ Clebsch--Gordan library}. It described the same
reduction of the bracket, Eq.~(\ref{eq:hob}), with the isofactors read from
the external vector-coherent-state library \Su{}~\cite{Bahri_2004}. The
present version replaces it rather than appearing separately, so that one
entry carries the program line.

That route is not discarded here: it is one of the three compared throughout,
in Sec.~\ref{sec:vcs} and Tables~\ref{tab:speed}--\ref{tab:domain}. The
comparison does, however, correct one claim of the earlier version, which
tested the mid-$L$ block $L=E/2$ at $E\le12$ and reported agreement with the
classical brackets ``to machine precision ($\sim\!10^{-13}$)'' without
qualification, ascribing the residual at $E=12$ to the double-precision floor
of the classical reference's binomial tables alone. Measured over
\emph{every} $(E,L)$ block up to $E=50$, the library route's own
orthonormality error is $4\times10^{-11}$ at $E=12$, $7\times10^{-3}$ at
$E=26$ and of order unity by $E\simeq29$ (Table~\ref{tab:accuracy}), and the
library returns non-finite values for every block with $E+L\ge60$
(Eq.~(\ref{eq:vcslimit})). The earlier claim therefore holds on the
best-conditioned blocks of the shells on which it was measured, and does not
extend beyond them; part of the $E=12$ residual is the isofactors, not the
reference. Nothing else in that version is withdrawn: its calling protocol,
its convention phase and its treatment of the scalar $(0,0)$ factor are
Sec.~\ref{sec:vcs} and Eq.~(\ref{eq:phase}) here.

\section{Introduction}
Harmonic-oscillator transformation brackets (HOBs), also called
Talmi--Moshinsky brackets, express a product of two
single-particle HO states in one set of coordinates as a
linear combination of products in a rotated set
of coordinates~\cite{Moshinsky,BuckMerchant,Kamuntavicius}. They are the
elementary ingredient in the construction of translationally invariant
few-body and \emph{ab initio} no-core shell-model matrix elements, where the
transformation between single-particle and relative--center-of-mass
coordinates is performed. Because realistic calculations
require HOBs for many two-oscillator blocks $(E,L)$ and a range of the
arbitrary mass-ratio parameter $d$, a compact and numerically
robust evaluation scheme always has value.

We recently \cite{KALINAUSKAS2025123033}
reformulated the general HOB entirely in the language of the $\su{SU}(3)$ group, reducing the bracket to a single sum over the small Wigner
$d$-functions of the
Talmi--Moshinsky transformation weighted by $\su{SU}(3)\supset\su{SO}(3)$
isofactors of the chain $\su{U}(6)\supset\su{U}(3)\times\su{U}(2)$.
The isofactors are $d$-independent, so once tabulated for a given $(E,L)$ they
serve for every value of $d$; this separation is the source of the method's
efficiency.

This paper is about how those isofactors are best obtained. The construction
adopted here uses nothing beyond the chain itself. The $\su{U}(2)$ factor is
generated by the pseudo-spin bilinears, whose raising operator is an
$\su{SO}(3)$ scalar; since the $\su{U}(3)$ and $\su{U}(2)$ labels inside $[E]$
of $\su{U}(6)$ are complementary, classifying that pseudo-spin at fixed
$(E,L)$ \emph{is} the $\su{SU}(3)\supset\su{SO}(3)$ classification. The
highest-weight states are then the null space of the raising operator in one
block, and the rest of each multiplet follows by lowering. No $\su{SU}(3)$
Clebsch--Gordan machinery enters at any stage: no weight recursions, no
$K$-matrix orthonormalization, and no outer-multiplicity resolution.

To place this on a firm footing, we compare it against two independent routes
to the same numbers: the identical SU(3) scheme with the isofactors instead
supplied by the general-purpose Clebsch--Gordan library
\Su{}~\cite{Bahri_2004}, which computes them by the vector-coherent-state (VCS)
method, and the classical Talmi--Moshinsky sum
\cite{Kamuntavicius,BuckMerchant}. \Su{} is used here as the library
comparator because it is the one that produces $\su{SU}(3)\supset\su{SO}(3)$
coefficients in the form the bracket needs; more recent SU(3) recoupling
machinery, in particular \code{SU3lib}~\cite{SU3lib} and the coefficient
engine inside \code{LSU3shell}~\cite{LSU3shell}, targets the canonical basis
and large-scale symmetry-adapted bases rather than this reduction, and is not
a drop-in alternative for the isofactor step. The brackets themselves are
consumed downstream by translationally invariant no-core shell-model
calculations~\cite{NCSM,Kamuntavicius}, which is the setting that fixes the
range of $E$ of interest here. The comparison is carried out block by block over the whole
range $E\le50$ and covers both cost and accuracy. It shows that the SU(3)
scheme is worth two to three orders of magnitude over element-wise evaluation
of the classical formula; that within that scheme the ladder construction is a
further factor of six to eight faster than the library route, and several
orders of magnitude more accurate; and that each of the two alternatives has
a ceiling in $E$ --- of accuracy for the library route, of cost for the
classical one --- that the ladder construction does not.

The remainder of this paper is organized as follows. Section~\ref{sec:theory}
recalls the SU(3)-basis expression for the HOB. Section~\ref{sec:iso} is the
core of the present contribution: it sets out the pseudo-spin ladder
construction on which the package is built, then the vector-coherent-state
route by which \Su{} obtains the same coefficients and which serves here as
an independent check, then how the latter are extracted in practice,
the convention phase relating the two libraries, and the treatment of the
scalar oscillator factor. Section~\ref{sec:prog} describes the
implementation and Sec.~\ref{sec:test} the validation, including a
cost comparison of the available routes to the same brackets.
Section~\ref{sec:cfp} carries that comparison into an application, the
three-particle coefficients of fractional parentage.
Section~\ref{sec:summary} concludes.

\section{Harmonic-oscillator brackets in the SU(3) basis}
\label{sec:theory}
We use the two-oscillator coupled states
$\ket{e_1 l_1, e_2 l_2 : L M}$, where $e_i$ is the number of oscillator
quanta and $l_i$ the orbital angular momentum of oscillator $i$, coupled to total
angular momentum $L$. The total number of quanta $E=e_1+e_2$ is conserved by the transformation,
so the HOB is block diagonal in $E$ and in $L$. Following \cite{KALINAUSKAS2025123033},
the general Talmi--Moshinsky bracket is
\begin{equation}
\begin{split}
  \big\langle e_1 l_1, e_2 l_2 : L \big|\, T(d)\, \big|
             e_1' l_1', e_2' l_2' : L \big\rangle
  = \sum_{J} \Delta_0^{\,E_2}\, & d^{\,J}_{M,M'}(d)
    \sum_{\alpha=1}^{\alpha_0}
      \iso{l_1}{l_2}{J}{\alpha}{e_1 e_2 L}\, \\
    &\times
      \iso{l_1'}{l_2'}{J}{\alpha}{e_1' e_2' L},
\end{split}
  \label{eq:hob}
\end{equation}
with $E=e_1+e_2=e_1'+e_2'$, $M=(e_1-e_2)/2$, $M'=(e_1'-e_2')/2$,
$E_2=E/2-J$, and $\Delta_0=\det(\mathrm{TM})=-1$ the determinant of the
Talmi--Moshinsky transformation [Eq.~(\ref{eq:tm})]. The label $J$ runs over
$J=E/2,\,E/2-1,\dots$ and fixes the $\su{SU}(3)$ irrep
$\lm=(2J,\,(E-2J)/2)$ contained in the coupled product $(e_1,0)\times(e_2,0)$.
The quantity $d^{J}_{M,M'}(d)$ is the small Wigner $d$-function of the
reflection-containing Talmi--Moshinsky matrix
\begin{equation}
  \mathrm{TM}=\begin{pmatrix}\cos\theta & \sin\theta\\[2pt]
                             \sin\theta & -\cos\theta\end{pmatrix},
  \qquad
  \det(\mathrm{TM})=-\cos^{2}\theta-\sin^{2}\theta=-1,
  \label{eq:tm}
\end{equation}
with $\cos\theta=\sqrt{d/(1+d)}$ and $\sin\theta=\sqrt{1/(1+d)}$. The lower-right
entry $-\cos\theta$ makes $\mathrm{TM}$ a reflection rather than a proper
rotation, so its determinant is the explicit value
$\Delta_0=\det(\mathrm{TM})=-1$ that enters the phase $\Delta_0^{E_2}$ of
Eq.~(\ref{eq:hob}). Then $d^{J}_{M,M'}(d)$ is the ordinary Wigner
$d^{J}_{M,M'}(\theta)$ at this reflection-augmented angle, implemented in the
routine \code{dtm} in a numerically stable logarithmic form.

The inner objects $\iso{l_1}{l_2}{J}{\alpha}{e_1 e_2 L}$ are the
$\su{SU}(3)\supset\su{SO}(3)$ isofactors (SO(3)-reduced Wigner coefficients)
for coupling the two symmetric irreps $(e_1,0)$ and $(e_2,0)$, with factor
SO(3) labels $l_1$ and $l_2$, to the resultant $\lm$ with total orbital
angular momentum $L$ in the $\alpha$-th occurrence. The \emph{inner}
multiplicity $\alpha_0$ --- the number of times $L$ appears within $\lm$, not
to be confused with the outer multiplicity $\rho$ of an irrep in a coupling,
which is $1$ throughout for the symmetric couplings needed here --- is given by
the Racah formula
\begin{equation}
\alpha_0 = \left[  \frac{\lambda+\mu-L+2}{2}\right]  _{\geq
0}-\left[  \frac{\lambda-L+1}{2}\right]  _{\geq0}-\left[  \frac{\mu-L+1}%
{2}\right]  _{\geq0}
\label{eq:mult}
\end{equation}
coded in the function \code{alpha\_mult} and cross-checked against the
number of isofactors returned by the coupling library.

For a fixed $(E,L)$ block Eq.~(\ref{eq:hob}) is equivalently a single matrix
product. Collecting the isofactors into the (state)$\times$(SU(3)-label) matrix
$C$ (rows $(e_i,l_i)$, columns $(J,\alpha,M)$) and the reflection $d$-functions
into $\mathcal{D}(d)$---block diagonal in $J$ and in the multiplicity $\alpha$,
with entries $\Delta_0^{E_2}\,d^{J}_{M,M'}(d)$---the bracket matrix of the whole
block is
\begin{equation}
  H(d) = C\,\mathcal{D}(d)\,C^{T},
  \label{eq:matrix}
\end{equation}
which is the form assembled by \code{hob\_eval}; evaluating Eq.~(\ref{eq:hob})
entry by entry (as in a direct element-wise transcription of the sum) gives
identical values. Because $C$ is independent of $d$, it is built once per
$(E,L)$---the isofactor towers---and reused for every $d$ and every matrix
element; this ``prepare once, evaluate many'' factorization is the origin of
the speed advantage over an element-wise evaluation that recomputes the
isofactors for each bracket.

Because Eq.~(\ref{eq:hob}) contracts the isofactors of the bra and ket as an
outer product $V V^{T}$ summed over the multiplicity index $\alpha$, the
assembled bracket is invariant under any orthogonal transformation of the
degenerate $\alpha$ basis. Consequently, the library's ordering/rotation of the
degenerate isofactors is irrelevant, and the orthonormality
$\mathrm{HOB}\cdot\mathrm{HOB}^{T}=I$ holds regardless of the
particular $\alpha$-basis convention adopted by a given coupling library. This
observation is central to the interface described next.

\section{The \texorpdfstring{$\su{SU}(3)\supset\su{SO}(3)$}{SU(3) > SO(3)} isofactors}
\label{sec:iso}

\subsection{The pseudo-spin \texorpdfstring{$\su{U}(2)$}{U(2)} ladder construction}
\label{sec:raising}
The isofactors of Eq.~(\ref{eq:hob}) are fixed by the group chain alone, so
they may be obtained in more than one way. The construction used here, and in
the original implementation~\cite{KALINAUSKAS2025123033}, calls no coupling
library at all: it generates them from the $\su{U}(2)$ factor of the chain
itself. This is the route the rest of the paper measures the alternatives
against.

The $\su{U}(2)$ factor of $\su{U}(6)\supset\su{U}(3)\times\su{U}(2)$ is
generated by the bilinears $\sum_{\mu}a^{\dagger}_{(t)\mu}a_{(t')\mu}$, which
in angular-momentum form read
\begin{equation}
  \hat J_{+}=\sum_{\mu}a^{\dagger}_{(1)\mu}a_{(2)\mu},
  \qquad
  \hat J_{-}=\hat J_{+}^{\dagger},
  \qquad
  \hat J_{z}=\tfrac12\bigl(\hat N_1-\hat N_2\bigr),
  \label{eq:su2gen}
\end{equation}
and satisfy the $\su{SU}(2)$ algebra. Three properties make this
\emph{pseudo-spin} the natural computational tool.

First, $\hat J_{\pm,z}$ commute with the total orbital angular momentum: the
generators of $\su{SO}(3)$ are diagonal in the coordinate index $t$, while
\eqref{eq:su2gen} is diagonal in the spherical index $\mu$. Hence $\hat J_{+}$
is an $\su{SO}(3)$ scalar, acts within a subspace of fixed $(E,L,M_L)$, and by
the Wigner--Eckart theorem its matrix does not depend on $M_L$ --- so the
construction may be carried out at $M_L=L$ once and for all.

Second, $\hat J_{z}$ is diagonal in the coupled basis, with eigenvalue
$M=\tfrac12(e_1-e_2)$. At fixed $(E,L)$ the basis therefore splits into blocks
of fixed $e_1$, and $\hat J_{+}$ moves one quantum from the second oscillator to
the first, mapping block $M$ to block $M+1$.

Third --- and this is what makes the method work --- the $\su{U}(3)$ and
$\su{U}(2)$ irrep labels inside $[E]$ of $\su{U}(6)$ coincide, so
\emph{classifying the pseudo-spin $\su{SU}(2)$ at fixed $(E,L)$ is already the
$\su{SU}(3)\supset\su{SO}(3)$ classification}: a pseudo-spin multiplet of
length $2J+1$ is one $\su{SU}(3)$ irrep $\lm=(2J,\tfrac E2-J)$ contributing one
state to every block $|M|\le J$, and the number $\alpha_0$ of degenerate
multiplets of the same $J$ is exactly the inner multiplicity of $L$ in that
irrep. The isofactor is then nothing but the coordinate vector of the
simultaneous eigenstate of $\hat{\bm J}^2$ and $\hat J_z$ in the block basis.
The whole problem thus reduces to diagonalizing the pseudo-spin $\su{SU}(2)$
inside each $(E,L)$ space, and no $\su{SU}(3)$ machinery --- weights,
$K$-matrices, or the Draayer--Akiyama and Bahri--Rowe--Draayer algorithms on
which \Su{} itself rests --- is needed at all.

The matrix elements of $\hat J_{+}$ are built from those of a single creation
operator, which in the coupled single-oscillator basis are
\begin{equation}
  \bigl(e{+}1,\,l{+}1\big\|\mathbf a^{\dagger}\big\|e\,l\bigr)
   =\sqrt{\frac{(l+1)(e+l+3)}{2l+3}},
  \qquad
  \bigl(e{+}1,\,l{-}1\big\|\mathbf a^{\dagger}\big\|e\,l\bigr)
   =\sqrt{\frac{l\,(e-l+2)}{2l-1}},
  \label{eq:redme}
\end{equation}
in the convention
$\me{n'l'm'}{a^{\dagger}_{\mu}}{nlm}
 =\braket{l\,m\,1\,\mu}{l'm'}\,\bigl(\cdot\big\|\mathbf a^{\dagger}\big\|\cdot\bigr)$.
Both branches are taken with a positive sign, which selects the radial phase
convention of the reference bracket codes~\cite{Kamuntavicius,BuckMerchant};
with the alternative $(-1)^{n}$ convention, the $l-1$ branch changes sign, and
every bracket acquires a factor $(-1)^{n_1+n_2+n_1'+n_2'}$. This is the same
kind of convention bookkeeping that reappears, in a different guise, as the
\Su{} phase of Sec.~\ref{sec:phase}.

The block structure induced by $\hat J_{z}$ is the working object throughout.
At fixed $(E,L)$, the coupled basis splits into blocks of fixed $e_1$, the block
$e_1$ being spanned by the pairs
\begin{equation}
  l_t\in\{e_t,\,e_t-2,\dots,1\ \text{or}\ 0\},
  \qquad |l_1-l_2|\le L\le l_1+l_2,
  \label{eq:blockbasis}
\end{equation}
with $e_2=E-e_1$; write $n_{e_1}$ for their number. The raising operator moves
one quantum from the second oscillator to the first, so it is a rectangular map
between adjacent blocks,
\begin{equation}
  \hat J_{+}^{(e_1)}:\ \text{block }e_1\longrightarrow\text{block }e_1{+}1,
  \qquad \dim = n_{e_1+1}\times n_{e_1},
  \label{eq:jpblock}
\end{equation}
and since every matrix element is real, the lowering operator needs no separate
construction: $\hat J_{-}^{(e_1+1)}=(\hat J_{+}^{(e_1)})^{\mathsf T}$.

Its matrix elements in the coupled basis follow from Eq.~(\ref{eq:redme}) by
applying the Wigner--Eckart theorem twice and summing over magnetic quantum
numbers,
\begin{equation}
\begin{split}
  \bigl(\hat J_{+}\bigr)^{(e_1)}_{(l_1'l_2'),(l_1l_2)}
  =\ &\bigl(e_1{+}1\,l_1'\big\|a^{\dagger}\big\|e_1 l_1\bigr)
      \bigl(e_2\,l_2\big\|a^{\dagger}\big\|e_2{-}1\,l_2'\bigr)\\
     &\times\sum_{m_1}\sum_{\mu=-1}^{1}
      \braket{l_1'\,m_1{+}\mu,\;l_2'\,L{-}m_1{-}\mu}{L\,L}\,
      \braket{l_1 m_1,\,l_2\,L{-}m_1}{L\,L}\\
     &\qquad\times
      \braket{l_1 m_1\,1\mu}{l_1'\,m_1{+}\mu}\,
      \braket{l_2'\,L{-}m_1{-}\mu,\;1\mu}{l_2\,L{-}m_1},
\end{split}
  \label{eq:jpelem}
\end{equation}
non-zero only for $l_1'=l_1\pm1$ and $l_2'=l_2\pm1$. The explicit sum has at
most $3(2l_1+1)$ terms and uses nothing but standard Condon--Shortley
$\su{SU}(2)$ Clebsch--Gordan coefficients. It could be condensed into a single
$6j$ symbol, but there is no reason to: assembling every $\hat J_{+}$ block is
a negligible fraction of the run time, and the explicit form sidesteps the
phase-convention pitfalls that a $6j$ recoupling would introduce --- which, as
Sec.~\ref{sec:cost} shows, is exactly where the classical route is most
fragile.

\emph{Highest weights.} A multiplet of pseudo-spin $J$ has its highest weight
in the block $M=J$, where $\hat J_{+}$ annihilates it. Conversely, any
vector of block $M$ annihilated by $\hat J_{+}$ heads a multiplet with $J=M$,
since a state of $J'>M$ inside the block has
$\|\hat J_{+}\psi\|^{2}=J'(J'+1)-M(M+1)>0$. Hence
\begin{equation}
  \operatorname{span}\bigl\{\ket{EJ\alpha L;M{=}J}\bigr\}_{\alpha=1}^{\alpha_0}
  =\ker \hat J_{+}^{(e_1^{\rm hw})},
  \qquad
  \alpha_0=\dim\ker \hat J_{+}^{(e_1^{\rm hw})}.
  \label{eq:kernel}
\end{equation}
The kernel is obtained from the spectral decomposition of the small, dense,
symmetric positive-semidefinite Gram matrix
\begin{equation}
  G=\bigl(\hat J_{+}^{(e_1^{\rm hw})}\bigr)^{\mathsf T}\hat J_{+}^{(e_1^{\rm hw})}
   =\hat J_{-}\hat J_{+}\big|_{\text{block }M}
   =\hat{\bm J}^{2}-M(M+1)\big|_{\text{block }M},
  \label{eq:gram}
\end{equation}
whose exact eigenvalues are $g_{J'}=J'(J'+1)-M(M+1)$. The null space is
therefore separated from the rest of the spectrum by a gap
$\min_{J'>M}g_{J'}=2(M+1)\ge 2$ that is \emph{independent of $E$ and $L$}, so the rank decision rests on a
\emph{relative} gap, since $\|G\|$ itself grows like $E^{2}$. That ratio is
also known exactly: the smallest non-zero eigenvalue is $2(J+1)$ and the
largest is $\tfrac E2(\tfrac E2+1)-J(J+1)$, so the worst case over a shell is
$J=0$, giving $2/[\tfrac E2(\tfrac E2+1)]$. At $E=50$ this is
$3.1\times10^{-3}$ --- still five orders above the $10^{-8}$ threshold used and
thirteen above double-precision round-off, so the null space is never in doubt.
Blocks with $M=E/2$, or whose neighbor is empty, are entirely highest weight
and need no diagonalization at all.

\emph{Lowering.} The remaining members of each multiplet follow from the
standard ladder relation
\begin{equation}
  \ket{EJ\alpha L;M{-}1}
  =\frac{\hat J_{-}\ket{EJ\alpha L;M}}{\sqrt{J(J+1)-M(M-1)}},
  \label{eq:lowering}
\end{equation}
applied down to $M=-J$. Reaching the lower weights this way, rather than by
solving an independent problem in each block, is what fixes one consistent
labeling $\alpha$ across the whole tower; solved separately, each $M$ would
return an arbitrary basis of its own degenerate space and the bra and ket
isofactors of Eq.~(\ref{eq:hob}) would not refer to the same $\alpha$.

The computed $\alpha_0$ is checked against the Racah inner multiplicity,
Eq.~(\ref{eq:mult}), for every $(E,J,L)$. This is a strong end-to-end test
rather than a formality: a wrong sign or weight anywhere in $\hat J_{+}$
destroys the exact $\su{SU}(2)$ spectrum $J'(J'+1)-M(M+1)$ and with it the
kernel dimensions.

Collecting the steps, the whole construction for one $(E,L)$ is the following,
and it is worth setting out in full because every step is elementary:

\begin{enumerate}\setlength{\itemsep}{3pt}
\item enumerate the block bases (\ref{eq:blockbasis}) for $e_1=0,\dots,E$;
\item build the rectangular $\hat J_{+}$ matrices (\ref{eq:jpelem}) between
      adjacent blocks;
\item for $2J=E,E-2,\dots$ take the kernel (\ref{eq:kernel}) in the block
      $M=J$, giving the highest-weight matrix $V^{J}_{J}$, check $\alpha_0$
      against the Racah count (\ref{eq:mult}), and lower with
      (\ref{eq:lowering}) to obtain $V^{J}_{M}$ for every $M$;
\item for any requested $d$, assemble the bracket as
      $\sum_{J}(-1)^{E/2-J}\,d^{J}_{M,M'}(d)\;
       V^{J}_{M}\bigl(V^{J}_{M'}\bigr)^{\mathsf T}$ over block pairs.
\end{enumerate}

Steps 1--3 involve no $d$ at all. This is where the ``prepare once, evaluate
many'' structure of Sec.~\ref{sec:prog} comes from, and it is a property of the
group chain rather than of the implementation: the isofactors label states of
$\su{U}(6)\supset\su{U}(3)\times\su{U}(2)$, while $d$ enters only through the
Wigner function of the coordinate transformation.

The cost of the construction is correspondingly small. With block dimensions
$n_M$ satisfying $\sum_M n_M=O(E^{2})$ at fixed $L$, steps 1--3 require $O(E)$
diagonalizations of symmetric matrices of size $O(E)$ and $O(E^{2})$ small
matrix products. Nothing in it grows like an $\su{SU}(3)$ recoupling: the
measured time for all $L$ blocks at $E=10$ is a few milliseconds, and the
whole $E=50$ shell --- $51$ blocks, the largest $2907\times2907$ --- is
prepared in $5.9$~s (Table~\ref{tab:speed}).

One factor in step 4 is easy to lose and worth recording, since omitting it is
silent. The $\su{U}(2)$ irrep $[E_1E_2]$ is $(\det)^{E_2}\otimes
\mathrm{Sym}^{2J}$, and the reflection part of the Talmi--Moshinsky matrix has
determinant $-1$, so the sum carries $\Delta_0^{E_2}=(-1)^{E/2-J}$. Its effect
is visible already in the simplest non-trivial block, $E=2$, $L=1$: the single
state $\ket{1\,1,1\,1{:}1}\propto(\mathbf a^{\dagger}_{(1)}\times
\mathbf a^{\dagger}_{(2)})\ket{0}$ has $J=0$ and $d^{0}_{00}=1$, yet the exact
bracket is $-1$ for every $d$, because
$\mathbf r_1\times\mathbf r_2\mapsto-\mathbf r_1'\times\mathbf r_2'$ under an
orthogonal transformation of determinant $-1$.

Finally, two choices in this construction are arbitrary --- the orthonormal
basis taken in each kernel, and the sign of each kernel vector. Neither
affects any bracket. The lowering map is linear and common to all $\alpha$, so
$\tilde V^{J}_{J}=V^{J}_{J}O$ with $O\in O(\alpha_0)$ implies
$\tilde V^{J}_{M}=V^{J}_{M}O$ for every $M$, and the only combination entering
Eq.~(\ref{eq:hob}) is the $\alpha$-summed dyadic
$V^{J}_{M}(V^{J}_{M'})^{\mathsf T}$, which is invariant. That dyadic is the
matrix of the orthogonal projector onto the $J$-multiplet subspace, transported
between blocks by the ladder operators, and is basis independent. The
isofactors generated here may therefore differ from tabulated
Bahri--Rowe--Draayer values by an orthogonal $\alpha$-mixing and by signs
whenever $\alpha_0>1$, while every physical bracket agrees --- which is exactly
what the comparison of Sec.~\ref{sec:test} confirms, and what makes it a
meaningful test rather than a tautology.

\subsection{How \texorpdfstring{\Su{}}{Su3cgvcs} computes them: the VCS route}
\label{sec:vcs}
The library solves a considerably more general problem than the one posed in
Sec.~\ref{sec:raising}, and it is worth setting the two side by side, since
that is what the cost comparison of Sec.~\ref{sec:cost} ultimately measures.

\Su{}~\cite{Bahri_2004} computes the Clebsch--Gordan coefficients for an
arbitrary coupling $\lm_1\times\lm_2\to\lm_3$ in either the canonical
$\su{SU}(3)\supset\su{SU}(2)\times\su{U}(1)$ or the physical
$\su{SU}(3)\supset\su{SO}(3)$ basis, by the vector-coherent-state method of
Rowe and Bahri. The route to an $\su{SO}(3)$-reduced coefficient has three
ingredients.

First, the coupling is solved where it is easiest. The routine \code{cgu3hw}
returns the $\su{SU}(3)\supset\su{U}(1)\times\su{SU}(2)$ coefficients for which
the \emph{resultant} irrep carries its highest weight; in VCS theory these
follow from a comparatively simple recursion, and they seed everything else.

Second --- and this is the step with no counterpart in the ladder construction
--- the $\su{SO}(3)$ states of an $\su{SU}(3)$ irrep must be orthonormalized.
In Elliott's labeling~\cite{Elliott} the intrinsic label $K$ is not a good
quantum number
when the inner multiplicity exceeds one, so states of the same $L$ but
different $K$ are \emph{not} orthogonal. Their overlap matrix $S^{L}(K,K')$ is
built in closed form by \code{kmat} and then diagonalized numerically
(Householder reduction followed by a QL iteration); the orthonormalizing
matrix $\bar K^{\lm}_{K\alpha}(L)$ is its inverse square root. It is this
$\bar K$ that carries the multiplicity label $\alpha$, and it must be
constructed for each of the three irreps entering a coupling.

Third, the $\su{SO}(3)$-reduced coefficient is assembled in \code{cgu3o3} by
transforming the highest-weight coefficients from the $\su{SU}(2)$ to the
$\su{SO}(3)$ chain and contracting with the three $\bar K$ matrices,
schematically
\begin{equation}
\begin{split}
  \bigl\langle \lm_1\alpha_1 L_1;\lm_2\alpha_2 L_2 \big\|
               \rho\,\lm_3\alpha_3 L_3 \bigr\rangle
  \;\sim\;
  &\sum_{M_1M_2}\sum_{K_1K_2K_3}\sum_{jIN}
   \bar K_{K_1\alpha_1}\bar K_{K_2\alpha_2}\bar K_{K_3\alpha_3}\\[-2pt]
  &\qquad\times\;
   \langle K L M | j I N\rangle\;
   \bigl\langle\cdot\big\|\rho\,\lm_3\,0\,s_3\bigr\rangle,
\end{split}
  \label{eq:vcs}
\end{equation}
with $\langle K L M|jIN\rangle$ the $\su{SO}(3)$--$\su{SU}(2)$ highest-weight
overlaps of \code{ovrlap}, and a final Gram--Schmidt pass (\code{orthon})
fixing the remaining freedom when the outer multiplicity $\rho$ exceeds one.

Applied to our problem, this machinery is doing more work than the problem
requires, in three specific ways. The couplings we need are of two
\emph{symmetric} irreps, $(e_1,0)\times(e_2,0)$, for which the outer
multiplicity is always $\rho=1$, so the entire $\rho$ apparatus --- and the
Gram--Schmidt pass that resolves it --- is redundant here; this is why the
isofactor is read from the single element \code{dcgu3o3(1,1,$\alpha$,1)} in
Sec.~\ref{sec:protocol}. The $\bar K$ matrices must nevertheless be built and
inverted for all three irreps of every coupling. And the coefficients are
produced one $(L_1,L_2,L_3)$ triple at a time, whereas the ladder construction
creates a complete $J$-multiplet across every block $M$ in a single pass.

The ladder route of Sec.~\ref{sec:raising} avoids all three. Because the
$\su{U}(3)$ and $\su{U}(2)$ labels inside $[E]$ of $\su{U}(6)$ are
complementary, the isofactors emerge already orthonormal, as eigenvectors of a
symmetric Gram matrix with an exactly known spectrum: no $K$-matrix, no inverse
square root, no outer multiplicity, and no $\su{SU}(3)$ recursion at all. That
is the structural reason the internal construction is an order of magnitude
faster in the preparation stage (Table~\ref{tab:speed}), and equally the reason the
library route is worth having: it is general, independently developed, and
therefore checks the specialized construction against a completely different
algorithm.

\subsection{Isofactors from \Su{}: calling protocol}
\label{sec:protocol}
The library \Su{}~\cite{Bahri_2004} computes SU(3) Clebsch--Gordan coefficients
in the physical SU(3)$\,\supset\,$SU(2)$\times$U(1) and
SU(3)$\,\supset\,$SO(3) bases by the vector-coherent-state method. The
SO(3)-reduced coefficient it returns for $(e_1,0)\times(e_2,0)\to\lm$ with
factor labels $l_1,l_2$ combining to $L$ is precisely the isofactor
$\iso{l_1}{l_2}{J}{\alpha}{e_1 e_2 L}$ of Eq.~(\ref{eq:hob}), up to the
convention phase, which is  discussed in Sec.~\ref{sec:phase}. The documented usage
protocol is followed in \code{hob\_prepare}:
\begin{enumerate}
\item once per run, \code{readfact} and \code{readtab} load the factorial /
      binomial and $I$-/$S$-function tables (this is done in \code{hob\_init});
\item for each factorization $e_1+e_2=E$ and target $\lm$, a call to
      \code{cgu3hw} with the factor irreps $(e_1,0)$, $(e_2,0)$ and the
      resultant $\lm$ builds the highest-weight seed coefficients;
\item a call to \code{cgu3o3} with the same irreps and the factor labels
      $l_1,l_2$ coupled to $L$ returns the SO(3)-reduced coefficients in the
      array \code{dcgu3o3}; the isofactor for multiplicity $\alpha$ is the
      element \code{dcgu3o3(1,1,$\alpha$,1)}.
\end{enumerate}
These values fill the tower arrays \code{S\%tw(it)\%V(ie)\%a(r,$\alpha$)}: one
column per multiplicity index $\alpha$, one row per $(l_1,l_2)$ state of the
block of fixed $e_1$.

\subsection{Convention phase}
\label{sec:phase}
\Su{} returns the SU(3)$\,\supset\,$SO(3) reduced Wigner coefficients in the
canonical (Draayer) VCS convention. This differs from the HO
(Moshinsky) isofactor convention adopted in Eq.~(\ref{eq:hob}) by a
state-dependent sign equal to the parity of the number of radial quanta of the
state,
\begin{equation}
  \iso{l_1}{l_2}{J}{\alpha}{e_1 e_2 L}\bigg|_{\text{Eq.\,(\ref{eq:hob})}}
  = (-1)^{\,n_1+n_2}\; \code{dcgu3o3}(1,1,\alpha,1),
  \qquad n_i = \frac{e_i - l_i}{2}.
  \label{eq:phase}
\end{equation}
As noted in Sec.~\ref{sec:theory}, the orthonormality of the assembled matrix
is insensitive to this phase. Attaching the sign $s_i=(-1)^{n_i}$ to state $i$
multiplies row $i$ of both $H$ and $H^{T}$, so in the product
$H H^{T}$ the two factors of $s_i$ meet on the diagonal, where
$s_i^2=1$; orthonormality is thus preserved regardless of the phase. The
individual bracket $H_{ij}$, however, carries a single factor
$s_i s_j=(-1)^{n_i+n_j}$ and does change---and it is the physical quantity.
Omitting the phase (\ref{eq:phase}) therefore leaves the matrix orthonormal but
gives every bracket the wrong sign $(-1)^{n_i+n_j}$ relative to the reference
value. Restoring the per-state phase
of Eq.~(\ref{eq:phase}) makes Eq.~(\ref{eq:hob}) reproduce the physical
Talmi--Moshinsky bracket exactly, in agreement with the general
Talmi--Moshinsky bracket code of Kamuntavi\v{c}ius et
al.~\cite{Kamuntavicius} to machine precision (Sec.~\ref{sec:test}).

\subsection{The scalar \texorpdfstring{$(0,0)$}{(0,0)} oscillator factor}
\label{sec:scalar}
When $e_1=0$ or $e_2=0$, one of the coupled irreps is the scalar $(0,0)$. The
coupling $(e_2,0)\times(0,0)\to(e_2,0)$ (or its mirror) is trivial: the block
contains a single $(l_1,l_2)$ state and the reduced coefficient equals unity,
up to the phase of Eq.~(\ref{eq:phase}). The general-purpose \code{cgu3o3}
routine, however, performs an $\su{SO}(3)$ $K$-matrix eigen-decomposition that
is degenerate for a scalar factor and is not guarded against it; because the
single-block driver distributed with \Su{} never encounters $e_i=0$, this case
was not exercised there. The full sweep over
$e_1=0,\dots,E$ needed here does reach it, so \code{hob\_prepare} sets the scalar-factor
isofactor analytically to $(-1)^{n_1+n_2}$ and skips the library call. This is
the only place where the interface departs from a literal call-through to
\Su{}, and it is exact.

\section{Implementation}
\label{sec:prog}
The computational core is the Fortran~2008 module \code{hob\_su3}. For a fixed
$(E,L)$ the derived type \code{hob\_solver\_t} holds the coupled basis and the
isofactor towers, one per allowed $J$, each storing the block-structured matrix
$V^{J}_{M}$ of isofactors indexed by $e_1$ and by $(\text{state},\alpha)$.
Building those towers is steps 1--3 of Sec.~\ref{sec:raising}: the block
enumeration, the rectangular $\hat J_{+}$ matrices, and the kernel-plus-lowering
pass. No $\su{SU}(3)$ coupling library is involved at any point --- the only
external coefficients used are ordinary $\su{SU}(2)$ Clebsch--Gordan
coefficients, from the bundled \code{WignerSymbol} module~\cite{WignerSymbol},
which enter the $\hat J_{+}$ matrix elements (\ref{eq:jpelem}) --- and the
only numerical linear algebra is the cyclic Jacobi diagonalization of the
small Gram matrices (\ref{eq:gram}).

The design point that matters physically is the separation already visible in
Eq.~(\ref{eq:hob}): tower construction is independent of $d$, so brackets for
any number of transformation parameters cost only small Wigner $d$-functions
and the matrix products of Eq.~(\ref{eq:matrix}). The workflow is ``prepare
once, evaluate many'', and it is what makes the SU(3)-scheme evaluation
competitive with a direct element-wise transcription of the bracket formula.

The isofactor source is a build-time choice. The default is the ladder
construction just described, which is self-contained; a second build path
replaces step 3 by calls to the external Clebsch--Gordan library
\Su{}~\cite{Bahri_2004}, obtained and licensed separately, using the protocol
of Sec.~\ref{sec:protocol}. Everything downstream of the towers --- Wigner
functions, assembly, the public interface --- is shared between the two. That
second path is not needed to use the package; it exists so that the isofactors
can be produced by a completely independent algorithm, which is what makes the
comparisons of Secs.~\ref{sec:cost} and \ref{sec:cfp} meaningful rather than
self-referential.

\subsection{Interface}
\label{sec:api}
The whole library is reached through six public procedures:
\begin{quote}\ttfamily\small
call hob\_init(Emax)\\
call hob\_build\_basis(E, L, b)\\
call hob\_prepare(E, L, S)\\
call hob\_eval(S, d, H)\\
val = hob\_bracket(S, e1,l1,e2,l2, e1p,l1p,e2p,l2p, d)\\
a0~~= alpha\_mult(E, twoJ, L)
\end{quote}
\code{hob\_init} loads the Wigner tables once. \code{hob\_prepare} performs
steps 1--3 of Sec.~\ref{sec:raising} for one block and stores the towers in a
\code{hob\_solver\_t}; it does not depend on $d$. \code{hob\_eval} returns the
complete bracket matrix of that block for a given $d$, and
\code{hob\_bracket} a single element. The intended usage pattern follows
directly:
\begin{quote}\ttfamily\small
call hob\_prepare(E, L, S)~~~~~! once per block\\
do i = 1, nd\\
~~call hob\_eval(S, d(i), H)~~~! any number of mass ratios\\
end do
\end{quote}

\subsection{Build, test and use}
\label{sec:use}
The package builds with any Fortran~2008 compiler and no external library:
\begin{quote}\ttfamily\small
make~~~~~~~~~~~~! library, validation driver, CLI\\
make check~~~~~~! validation against the stored reference\\
make bench~~~~~~! timing and accuracy drivers
\end{quote}
\code{make check} runs the driver of Sec.~\ref{sec:test}, which checks, for
every block up to $E_{\max}$ and several $d$: the multiplicity $\alpha_0$
against the Racah count; orthonormality
$\|\mathrm{HOB}\cdot\mathrm{HOB}^{T}-I\|$; the involution property
$\|\mathrm{HOB}^{2}-I\|$, which follows because $T(d)$ is a reflection;
element-by-element agreement with the classical Talmi--Moshinsky formula,
evaluated independently by \code{tmb\_kam}; and the S(3) relations carried by
the $d=\tfrac13$ brackets, $\hat P_{23}^{2}=\Id$ and
$(\hat P_{12}\hat P_{23})^{3}=\Id$. Of these, the last two --- the S(3)
relations and the element-wise comparison --- are the ones that catch a
spurious orthogonal rotation of a block, to which orthonormality is blind. The
involution is only partly independent: under a conjugation $H\to OHO^{T}$ it
gives $H^{2}\to OH^{2}O^{T}=I$ and is blind in exactly the same way, and since
$H$ is symmetric $\|H^{2}-I\|$ and $\|HH^{T}-I\|$ are the same statement up to
rounding, as the printed output below shows. It does catch a one-sided
$H\to OH$, which orthonormality does not. It compares the numerical output
with a stored reference; the wall-clock section it also
prints is excluded from the comparison, since it cannot reproduce across
machines. BLAS/LAPACK is required only by \code{make bench}. The
command-line front end \code{hob} prints a single bracket or a whole block for
given $(E,L,d)$.

\subsection{Parallelism}
\label{sec:par}
All timings reported here are single-core, which is the conservative choice
for the comparisons of Sec.~\ref{sec:cost} --- one of the comparators
\cite{HOTB2014} is distributed as a parallel code. The blocks of a shell are
independent: nothing is shared between different $(E,L)$, so a loop over $L$
(or over shells) is embarrassingly parallel and scales trivially with an
\code{OMP PARALLEL DO} around the driver loop. Within a block, the cost is
dominated by the tower construction and by dense matrix products, both of
which inherit the threading provided by the linked BLAS. We have not
included a threaded driver in the distribution, since the appropriate
granularity depends on how the calling code sweeps its model space.

\section{Validation}
\label{sec:test}
We have checked the brackets over all blocks $(E,L)$ with $E\le E_{\max}$,
$0\le L\le E$, for three representative mass-ratio parameters
$d\in\{1,2,\tfrac12\}$, using three independent measures: (i) the orthonormality error
$\max|\mathrm{HOB}\cdot\mathrm{HOB}^{T}-I|$, the error measure of
Ref.~\cite{KALINAUSKAS2025123033}; (ii) the maximum absolute deviation of every bracket from
the general Talmi--Moshinsky value of Kamuntavi\v{c}ius et
al.~\cite{Kamuntavicius}, evaluated independently by the reference module
\code{tmb\_kam}; and (iii) the multiplicity count of Eq.~(\ref{eq:mult}).

The reference module deserves a word, because it is what fixes the range over
which an absolute test is possible at all. It implements the compact
expression of Ref.~\cite{Kamuntavicius} directly, with the $6j$, $9j$ and
Clebsch--Gordan coefficients in the forms given by Varshalovich, Moskalev and
Khersonskii~\cite{Varshalovich} --- Racah's single sum for the $6j$
(their Sec.~9.2.1), the single sum over three $6j$ for the $9j$
(Sec.~10.2.4), and Racah's formula for the Clebsch--Gordan coefficient
(Sec.~8.2.1) together with its zero-projection closed form (Sec.~8.5.1),
which is the only one the bracket actually needs. The one substantive
departure from the published Fortran is that every factorial and double
factorial is held as a logarithm rather than as a table of binomial and
trinomial coefficients. Those tables are exact only while their entries stay
below $2^{53}$, which is what caps the distributed implementation near
$E\simeq12$; in logarithmic form the same expression is evaluated without
overflow to $E=50$ and beyond, so the comparison below is limited by the
SU(3)-scheme bracket rather than by its yardstick.
Representative output for $E_{\max}=8$ is
\begin{lstlisting}
   d    max|HOB.HOB^T - I|  max|HOB^2 - I|  max|SU3HOB - Kam01|
  1.00        1.934009E-13    1.936229E-13         5.251355E-14
  2.00        2.111644E-13    2.109424E-13         5.617729E-14
  0.50        1.716405E-13    1.716405E-13         4.340972E-14
alpha0 multiplicities agree with the Racah formula for all (E, J, L).
\end{lstlisting}
(In the driver's output, and in the tables below that quote it, the computed
SU(3)-basis bracket is labelled \code{SU3HOB} --- the quantity, after the
engine module \code{hob\_su3} --- rather than \code{SU3HOB-ladder}, the
package.) Crucially, the
agreement with the general Talmi--Moshinsky reference is an \emph{absolute}
test of the bracket values, so it validates not only orthonormality (which is
convention-insensitive) but also the convention phase of Eq.~(\ref{eq:phase}).

The agreement holds block by block: a companion driver \code{test\_hob\_highe}
evaluates the representative mid-$L$ blocks $L=E/2$ for $E=2,4,\dots,12$, listed
in Table~\ref{tab:highe}. Both the orthonormality error and the deviation from
the independent reference grow smoothly and slowly, from $10^{-15}$ at $E=2$ to
a few times $10^{-12}$ at $E=26$; nothing in the range shows the abrupt loss of
precision that the binomial-table implementation suffers past $E\simeq12$. The
two columns track one another to within an order of magnitude throughout, which
is what one expects if the residual in both is the ordinary rounding of the
SU(3)-scheme construction rather than a discrepancy between the two routes to
the bracket. Timings for the same construction are collected in
Table~\ref{tab:speed}, alongside the two alternative routes.

\begin{table}[t]
\caption{Block-by-block validation on the mid-$L$ blocks $L=E/2$ at $d=1$
(the $d=2,\tfrac12$ values differ only in the last digit): block dimension,
orthonormality error, and maximum deviation of every bracket from the
independent classical reference \code{tmb\_kam}, which evaluates the compact
expression of Kamuntavi\v{c}ius et al.~\cite{Kamuntavicius} with the
angular-momentum coefficients of Ref.~\cite{Varshalovich}. Both measures are
compared over the complete block matrix, not a sample.}
\label{tab:highe}
\centering
\begin{tabular}{r r r c c}
\toprule
$E$ & $L$ & dim & $\max|\mathrm{HOB}\cdot\mathrm{HOB}^{T}-I|$
    & $\max|\mathrm{SU3HOB}-\mathrm{Kam01}|$ \\
\midrule
 2 &  1 &   1 & $0.0\times10^{0\hphantom{-1}}$ & $1.3\times10^{-15}$ \\
 4 &  2 &   9 & $4.4\times10^{-15}$ & $1.3\times10^{-15}$ \\
 6 &  3 &   9 & $1.0\times10^{-14}$ & $2.4\times10^{-15}$ \\
 8 &  4 &  30 & $2.4\times10^{-14}$ & $4.8\times10^{-15}$ \\
10 &  5 &  30 & $5.2\times10^{-14}$ & $1.0\times10^{-14}$ \\
12 &  6 &  70 & $5.6\times10^{-14}$ & $2.8\times10^{-14}$ \\
14 &  7 &  70 & $1.4\times10^{-13}$ & $3.6\times10^{-14}$ \\
16 &  8 & 135 & $3.6\times10^{-13}$ & $8.4\times10^{-14}$ \\
18 &  9 & 135 & $2.8\times10^{-13}$ & $9.4\times10^{-14}$ \\
20 & 10 & 231 & $9.4\times10^{-13}$ & $3.2\times10^{-13}$ \\
22 & 11 & 231 & $2.1\times10^{-12}$ & $8.4\times10^{-13}$ \\
24 & 12 & 364 & $7.3\times10^{-12}$ & $2.2\times10^{-12}$ \\
26 & 13 & 364 & $7.2\times10^{-12}$ & $2.0\times10^{-12}$ \\
\bottomrule
\end{tabular}
\end{table}


\subsection{Three routes to the same brackets}
\label{sec:cost}

Everything in this section compares four independent ways of producing
identical numbers, so their costs and their errors can be set directly against
one another. Two of them build the isofactors of Eq.~(\ref{eq:hob}) and
assemble the bracket by Eq.~(\ref{eq:matrix}):
\begin{itemize}\setlength{\itemsep}{2pt}
\item \textbf{ladder} --- the pseudo-spin $\su{U}(2)$ construction of
  Sec.~\ref{sec:raising}, Ref.~\cite{KALINAUSKAS2025123033};
\item \textbf{VCS} --- the same brackets with the isofactors taken from
  \Su{}~\cite{Bahri_2004}, Sec.~\ref{sec:vcs}.
\end{itemize}
The third bypasses the SU(3) scheme entirely and evaluates every matrix
element from the classical Talmi--Moshinsky sum:
\begin{itemize}\setlength{\itemsep}{2pt}
\item \textbf{Kam01} --- the compact expression of Kamuntavi\v{c}ius et
  al.~\cite{Kamuntavicius}, Eq.~(26) of that paper, with the $6j$, $9j$ and
  Clebsch--Gordan coefficients of Varshalovich et al.~\cite{Varshalovich} and
  every factorial and double factorial held in logarithmic form
  (\code{tmb\_kam}, Sec.~\ref{sec:test}).
\end{itemize}
The logarithmic tabulation matters for what follows. Implementations that
tabulate the binomial and trinomial coefficients themselves --- among them the
distributed HOTB code~\cite{HOTB2014} and the variant using the optimized $6j$
of Tuzun et al.~\cite{Tuzun1998} --- hold entries that are exact only while
they remain below $2^{53}$, and their internal arrays as shipped stop at
$n=99$ and $n=100$ respectively. Both bounds are reached well before the
shells of interest here. Taking logarithms removes the ceiling outright and
lets a single classical implementation stand for the whole family, which is
why one column suffices below where earlier drafts of this comparison carried
two.
All timings are for one core (\code{gfortran}~14.2, \code{-O2}, single-threaded
BLAS), summed over every block of the shell, at $d=\tfrac13$.

\begin{table}[htb]
\centering
\caption{Time in seconds to build every bracket of one shell, summed over all
$L$. For the two SU(3)-scheme routes this is the isofactor construction plus
the evaluation; the classical route has no separable preparation stage. The
classical column was not carried beyond $E=32$: one shell there already costs
ten minutes, and the cost grows as $n^{3.9}$ (Eq.~(\ref{eq:growth})), so this
is a limit of patience rather than of the method. The VCS column is set in
parentheses from $E=30$ on, where that route no longer returns valid numbers
for every block (Table~\ref{tab:domain}); we did not run it at $E=32$.}
\label{tab:speed}
\begin{tabular}{rrrrr}
\toprule
$E$ & $n_{\max}$ & ladder & VCS & Kam01\\
\midrule
10 &   50 &  0.001 &  0.004 &    0.01\\
16 &  147 &  0.008 &  0.031 &    0.67\\
20 &  252 &  0.028 &  0.111 &    5.24\\
24 &  405 &  0.073 &  0.376 &   31.34\\
26 &  495 &  0.123 &  0.660 &   70.15\\
28 &  605 &  0.190 &  1.030 &  150.41\\
30 &  726 &  0.301 & (1.87) &  309.65\\
32 &  858 &  0.444 & ---    &  599.95\\
36 & 1183 &  1.013 & (7.31) &  ---   \\
40 & 1575 &  2.156 &(16.76) &  ---   \\
44 & 2040 &  4.541 &(35.18) &  ---   \\
50 & 2907 & 12.718 &(96.81) &  ---   \\
\bottomrule
\end{tabular}
\end{table}

Table~\ref{tab:speed} separates two effects that are easily conflated. The
SU(3) scheme itself is worth two to three orders of magnitude over the
classical evaluation --- at $E=30$, $0.30$~s against $310$~s, a factor $1030$ ---
because Eq.~(\ref{eq:matrix}) produces a whole block as one matrix product
while the classical route recomputes a nested sum for each of the $n^{2}$
entries independently. Within the SU(3) scheme, the choice of isofactor source
is worth a further factor of $6$--$8$, and that factor sits entirely in the
preparation stage: the evaluation times of the two routes agree to within
run-to-run noise, as they must, since once the towers exist both execute the
same arithmetic. Because the towers are $d$-independent, that penalty is
amortized away in any application that reuses a block for several mass ratios.

The classical cost is set by the number of independent element evaluations,
not by the arithmetic inside them: substituting the optimized $6j$ of Tuzun
et al.~\cite{Tuzun1998} for a self-contained one changes the total by a few
per cent in either direction, and the logarithmic tabulation used here costs
about a quarter more than raw binomial lookups while removing their range
limit entirely. None of these choices alters the order of magnitude.

The classical column of Table~\ref{tab:speed} stops at $E=32$ for a reason
worth quantifying, since it also fixes what can be said about the shells
beyond. The cost is polynomial in the block dimension, not geometric in $E$, so
it should be fitted in $n$. Doing so gives a remarkably stable exponent for the
classical route,
\begin{equation}
  t \propto n^{\,p},\qquad p\simeq3.9,
  \label{eq:growth}
\end{equation}
(a least-squares fit over the seven measured points $n=147\to858$ gives
$3.85$, and the pairwise exponents scatter between $3.77$ and $4.02$ without
trend), while the same fit applied to the ladder route gives
$p\simeq2.5$--$2.9$, rising slowly as the linear algebra begins to dominate.
Extrapolating Eq.~(\ref{eq:growth}) from the measured $600$~s at $n=858$ to
$n_{\max}=2907$ puts a single $E=50$ shell at
\begin{equation}
  t(E{=}50)\simeq600\times(2907/858)^{3.85}\approx6.6\times10^{4}\ \mathrm{s}
  \approx18\ \mathrm{h}
\end{equation}
and the whole tabulation to $E=50$ at a day or two rather than the
$13$~s the ladder construction needs for that shell.

Two remarks make that estimate firmer than a single extrapolation. First,
$n_{\max}$ is a proxy: the work in a shell is proportional to the number of
brackets it contains, $\sum_{L}n_{L}^{2}$, and the ratio of that to
$n_{\max}^{2}$ drifts as the number of blocks increases. Fitting the same timings
against the bracket count isolates the per-element cost and gives
\begin{equation}
  t\propto\Bigl(\textstyle\sum_{L}n_{L}^{2}\Bigr)^{q},\qquad q\simeq1.60,
  \label{eq:growth2}
\end{equation}
the excess over unity being the growth of the per-element cost itself, which
rises from $1.4\times10^{-6}$~s per bracket at $E=10$ to
$6.2\times10^{-5}$~s at $E=32$. Extrapolated from
$\sum_{L}n_{L}^{2}=9\,684\,033$ at $E=32$ to $175\,680\,531$ at $E=50$, this
gives $17$~h, consistent with Eq.~(\ref{eq:growth}). Second, the same figure
follows without any fit at all: the $E=50$ shell contains $175\,680\,531$
brackets, and at the measured $0.37$~ms per bracket at that $E$ that is
$6.5\times10^{4}$~s, or $18$~h. Three routes to the same number, one of them
free of extrapolation, is a good deal stronger than an exponent carried a
factor $3.4$ in $n$ beyond the data. The more robust statement
is the difference of exponents: the gap between the two routes widens like
$n^{1.2}$, independently of any wall-clock extrapolation. Either way, the
conclusion is unchanged --- this is not a limitation of the Talmi--Moshinsky
formula, which remains the accuracy reference of Sec.~\ref{sec:cfp}, but a
statement about evaluating $n^{2}$ independent nested sums when $n$ has reached
$2907$.

\begin{table}[htb]
\centering
\caption{Accuracy of the three routes. The second and third columns are
$\max|\mathrm{HOB}\cdot\mathrm{HOB}^{T}-I|$ over the shell, the error measure
of Ref.~\cite{KALINAUSKAS2025123033}; the last is the largest deviation of any
bracket of the shell from the ladder value. Dashes mark shells not reached by
the classical route; the VCS column is bracketed where that route has begun to
fail outright; we did not run it at $E=32$, denoted by a dash as in
Table~\ref{tab:speed}. The last column is a difference and not an error of either
route on its own: from $E\simeq26$ it is dominated by the ladder residual in
the second column, which is why the two track one another rather than
diverging.}
\label{tab:accuracy}
\begin{tabular}{rlll}
\toprule
$E$ & ladder & VCS & $|\text{lad}-\text{Kam01}|$\\
\midrule
 8 & $1.5\times10^{-14}$ & $2.2\times10^{-13}$ & $1.7\times10^{-14}$\\
12 & $6.4\times10^{-14}$ & $4.1\times10^{-11}$ & $4.2\times10^{-14}$\\
16 & $1.1\times10^{-13}$ & $8.7\times10^{-9}$  & $1.7\times10^{-13}$\\
20 & $1.2\times10^{-12}$ & $2.7\times10^{-6}$  & $6.9\times10^{-13}$\\
24 & $5.0\times10^{-12}$ & $9.0\times10^{-4}$  & $2.6\times10^{-12}$\\
26 & $6.9\times10^{-11}$ & $7.2\times10^{-3}$  & $2.0\times10^{-11}$\\
28 & $3.3\times10^{-11}$ & $5.3\times10^{-2}$  & $1.3\times10^{-11}$\\
29 & $4.3\times10^{-10}$ & $8.6\times10^{-1}$  & $1.2\times10^{-10}$\\
30 & $1.5\times10^{-10}$ & (fails)             & $7.9\times10^{-11}$\\
32 & $2.8\times10^{-10}$ & ---                 & $7.2\times10^{-11}$\\
36 & $2.0\times10^{-8}$  & (fails)             & ---\\
44 & $3.7\times10^{-7}$  & (fails)             & ---\\
50 & $1.8\times10^{-6}$  & (fails)             & ---\\
\bottomrule
\end{tabular}
\end{table}

Table~\ref{tab:accuracy} is the more consequential comparison, and it sharply separates
the three routes. One limitation of the measure should be stated first,
because it is the reason Sec.~\ref{sec:cfp} exists.
$\|\mathrm{HOB}\cdot\mathrm{HOB}^{T}-I\|$ is necessary but not sufficient: it
is exactly zero under any spurious \emph{orthogonal} transformation of a
block. This is the same invariance that makes the $\alpha$-basis choice
harmless in Sec.~\ref{sec:raising} and the convention phase harmless in
Sec.~\ref{sec:phase} --- benign there, but here it means the columns below
bound only the non-orthogonal part of the error. The integer test of
Sec.~\ref{sec:cfp}, which is sensitive to any error at all, is what closes
that gap.

The ladder construction degrades slowly and predictably, from $10^{-14}$ at
$E=8$ to $2\times10^{-6}$ at $E=50$ --- ordinary floating-point accumulation in
a calculation whose dimension has grown to $2907$. It is the only one of the
three that remains usable over the whole range.

It is worth saying where that accumulation actually happens, because the
natural guess is wrong. The obvious suspect is the lowering recursion, which
at $E=50$ is applied up to fifty times in succession; the obvious remedy is
to re-orthonormalize the $n\times\alpha_0$ matrix $V^{J}_{M}$ after each step,
which costs nothing since $\alpha_0$ is small. Neither survives measurement.
The columns of $V^{J}_{M}$ are still orthonormal to $3\times10^{-11}$ at the
bottom of the longest ladder at $E=50$, five orders better than the bracket
error they are supposed to explain, and inserting a modified Gram--Schmidt
pass after every lowering step changes
$\max|\mathrm{HOB}\cdot\mathrm{HOB}^{T}-I|$ at $E=50$ from
$1.832\times10^{-6}$ to $1.833\times10^{-6}$. Evaluating the Wigner
$d$-functions of Eq.~(\ref{eq:hob}) in quadruple precision changes it by as
little.

What does track the bracket error is the mutual orthogonality of \emph{different}
$J$ towers. Collecting, block by block, every tower vector that lives in that
block into one matrix $W$ and measuring $\max|W^{\mathsf T}W-I|$ gives
$1.4\times10^{-13}$ at $E=16$, $3.7\times10^{-10}$ at $E=30$ and
$5.6\times10^{-6}$ at $E=50$ --- the same numbers, shell for shell, as the
second column of Table~\ref{tab:accuracy}. Towers with different $J$ are
orthogonal by symmetry and are never orthogonalized against one another; each
is built independently from its own kernel, and the small rounding errors of
separate constructions do not cancel. That is the accuracy limit of the
method as implemented, and it is not addressed by any of the three remedies
usually proposed for a lowering recursion.

The VCS route degrades far faster: already $10^{-11}$ at $E=12$, $10^{-6}$ at
$E=20$, and $10^{-3}$ by $E=26$, some seven orders of magnitude worse than the
ladder at the same shell. The natural explanation is the $\bar K$ matrix of
Sec.~\ref{sec:vcs}, which is the inverse square root of the overlap matrix
$S^{L}(K,K')$ and would pass that matrix's conditioning on to every isofactor.
We instrumented \code{kmat} to record $\mathrm{cond}(S^{L})$, taking the worst
over every irrep $(\lambda,\mu)=(E-2k,k)$ and every $L$ of the shell:
\begin{center}\small
\begin{tabular}{rrrrrrrr}
\toprule
$E$ & 8 & 12 & 16 & 20 & 24 & 30 & 50\\
$\mathrm{cond}(S^{L})$ & $3.2$ & $4.7$ & $8.8$ & $15$ & $27$ & $1.3\times10^{2}$
 & $2.7\times10^{4}$\\
\bottomrule
\end{tabular}
\end{center}
It grows, but nowhere near fast enough to be the cause. At $E=24$, where the
VCS bracket error has already reached $9\times10^{-4}$,
$\mathrm{cond}(S^{L})=27$ and the amplification through an inverse square root
is $\sqrt{27}\approx5$ --- five orders short of what would be needed. Even at
$E=50$ the factor is only $\sim1.6\times10^{2}$. The conditioning of $S^{L}$ is
therefore not the mechanism, and we do not claim it is; the loss must come from
elsewhere in the VCS chain --- the highest-weight construction, the
outer-multiplicity resolution, or the orthonormalization that follows --- which
we have not isolated. What can be said is the structural contrast: the ladder
construction forms no inverse square root and no ill-conditioned intermediate
of any kind, since complementarity delivers orthonormal vectors directly from a
Gram matrix whose spectrum is known exactly in advance.

The classical route behaves quite differently from either of them. Its last column is
a \emph{difference} between two independent evaluations of the same bracket,
and it stays at the level of the ladder residual itself all the way to $E=32$,
tracking the second column rather than drifting away from it. Neither route is
therefore resolving the other's error: what the column measures from
$E\simeq26$ on is the ladder's own rounding, seen through an independent
yardstick. This is the outcome that the logarithmic tabulation buys. The
accuracy of the classical formula is governed by how its combinatorial factors
are held and not by the formula itself --- the mirror image of the speed
result, where the same choice made almost no difference at all. Evaluated with
raw binomial tables, the identical expression has lost every significant digit
by $E=30$ ($6.4\times10^{-1}$); evaluated in logarithms, it is still at
$7.9\times10^{-11}$. Table~\ref{tab:tabulation} gives both side by side.

\begin{table}[htb]
\centering
\caption{Domain of validity actually observed. The last row is the same
classical formula as the third, evaluated instead from tabulated binomial and
trinomial coefficients: the two limits quoted for it are the point at which
such tables cease to be exactly representable and the point at which the
arrays shipped with the published codes overflow. The $E\le32$ entry for
Kam01 is a limit of this study, set by cost (Eq.~(\ref{eq:growth})), not by
any observed failure.}
\label{tab:domain}
\begin{tabular}{lll}
\toprule
route & usable range & failure mode beyond it\\
\midrule
ladder & $E\le50$ (all tested) & none observed\\
VCS & $E+L<60$ & non-finite values returned\\
Kam01 (logarithmic) & $E\le32$ (all tested) & none observed\\
Kam01 (binomial tables) & $E\simeq12$; $E\le29$ & silent precision loss;
  table overflow\\
\bottomrule
\end{tabular}
\end{table}

One limit deserves to be stated precisely, because it is sharp and was not
anticipated. From $E=30$ the VCS route returns non-finite values, and the
affected blocks are exactly those with
\begin{equation}
  E+L\ge60 .
  \label{eq:vcslimit}
\end{equation}
At $E=30$ this is the single block $L=30$; at $E=40$ it is the twenty-one
blocks $L\ge20$. The boundary is reproducible in the package's own build
directory and is independent of the driver, so it is a property of the coupling
library as distributed rather than of the interface described here. The
obvious question is whether the ceiling is simply a declared array bound, as
it is in the binomial-table classical codes, and can be lifted by enlarging
it. We attempted that: the two table-loading routines of the library
declare their binomial ranges with different bounds ($n\le400$ in one,
$n\le128$ in the other), and raising the smaller to match does \emph{not}
extend the range. It makes matters worse, with blocks failing from $E=2$,
which indicates that the tabulated $I$- and $S$-functions are consistent only
with the shipped bound and would have to be regenerated rather than merely
redeclared. That is a change to the library's internals rather than to its
build parameters, and we have not pursued it; the limit reported here should
therefore be read as a property of \Su{} as distributed, and it remains open
whether the underlying VCS method carries it. Users of the library route should treat
Eq.~(\ref{eq:vcslimit}) as a hard ceiling; the ladder construction has no
corresponding limit and was verified over every block to $E=50$.

\section{Application: coefficients of fractional parentage for three particles}
\label{sec:cfp}

The brackets are rarely wanted for their own sake, so it is worth carrying the
comparison one step into an application that consumes them. The three-particle
coefficients of fractional parentage (CFPs) are a good test case: they are the
first link in the antisymmetrization chain of translationally invariant
few-body and no-core shell-model calculations
\cite{Kamuntavicius}, they are needed block by block exactly as the brackets
are, and the quantity they are built from is precisely a Talmi--Moshinsky
bracket.

For three equal-mass particles in the Jacobi basis $\ket{e_1l_1,e_2l_2:L}$,
with
\begin{equation}
  \boldsymbol\rho_1=\tfrac{1}{\sqrt2}(\mathbf r_1-\mathbf r_2),
  \qquad
  \boldsymbol\rho_2=\tfrac{1}{\sqrt6}(\mathbf r_1+\mathbf r_2-2\mathbf r_3),
  \label{eq:jacobi}
\end{equation}
the transposition of the two particles inside the pair is diagonal,
$\hat P_{12}=\hat\Pi_1$ with $\hat\Pi_1\ket{i}=(-1)^{l_1}\ket{i}$, while the
two transpositions involving the spectator are brackets at $d=\tfrac13$:
\begin{equation}
  \hat P_{23}=\hat H_{1/3},
  \qquad
  \hat P_{13}=\hat\Pi_2\,\hat H_{1/3}\,\hat\Pi_2,
  \qquad
  \hat\Pi_2\ket{i}=(-1)^{l_2}\ket{i}.
  \label{eq:transp}
\end{equation}
The orbital CFPs are the orthonormal eigenvectors of the class operator
\begin{equation}
  \hat\Lambda=\hat P_{13}+\hat P_{23},
  \label{eq:lambda}
\end{equation}
obtained in the standard way: $\hat\Lambda$ commutes with $\hat P_{12}$, so it
is block diagonal in the two $\hat\Pi_1$ parity sectors, and each sector is
handed to a symmetric eigensolver. The resulting eigenvectors carry the S(3)
symmetry labels, and this is the construction used in practice
\cite{Kamuntavicius}.

What makes it a sharp diagnostic here is that the answer is known in advance.
On the S(3) irreps $\hat P_{12}+\hat P_{13}+\hat P_{23}=3,0,-3$ for
$[3]$, $[21]$, $[111]$, so the eigenvalues of $\hat\Lambda$ can only be the
exact integers
\begin{equation}
  \lambda\in\{+2,\,-2,\,+1,\,-1\},
  \label{eq:exactint}
\end{equation}
with $+2$ on the totally symmetric $[3]$, $-2$ on the totally antisymmetric
$[111]$, and $\pm1$ on the mixed-symmetry $[21]$ pairs. Any departure of a
computed eigenvalue from the nearest of these four integers is error inherited
from the brackets that built $\hat\Lambda$ --- the diagonalization itself
contributes nothing at this level, since a symmetric eigensolver is backward
stable and the matrices are small. The quantity
\begin{equation}
  \varepsilon = \max_{\text{block}}\;\bigl|\lambda-\mathrm{nint}(\lambda)\bigr|
  \label{eq:interr}
\end{equation}
is therefore a direct read-out of bracket quality, on the scale of the physical
answer rather than of an abstract matrix norm. Table~\ref{tab:cfp} gives it for CFPs built
from each of the three bracket routes of Sec.~\ref{sec:cost}.

The classical column deserves one remark, because the naive implementation of
the same formula does not reach this far. Codes that tabulate the binomial and
trinomial coefficients directly hold entries that are exact only below
$2^{53}$, and their arrays as shipped stop at $n\simeq100$; the first bound
costs precision silently from $E\simeq12$, the second stops the calculation
outright near $E=29$. Enlarging the arrays addresses only the second, and
postpones rather than removes the first. Holding every factorial and double
factorial as a logarithm removes both at once, at the price of the few digits
the alternating sums lose to cancellation, and this is how \code{tmb\_kam} is
written. It is what makes the $\varepsilon$ column meaningful over the entire
tabulated range. Table~\ref{tab:tabulation} contrasts the two tabulations of
the identical expression, run through the identical driver: by $E=30$ the
raw-binomial form returns eigenvalues wrong in the first decimal, so no
symmetry classification is possible at all, while the logarithmic form is
still at $10^{-10}$.

\begin{table}[htb]
\centering
\caption{The same classical expression, the same driver, two tabulations of
its combinatorial factors. Column two is the largest deviation of any bracket
of the shell from the ladder value; column three is $\varepsilon$ of
Eq.~(\ref{eq:interr}) for the CFPs built from those brackets. The binomial
column is the \code{C6J} route of the published HOTB
implementation~\cite{Kamuntavicius,HOTB2014} with its internal array enlarged
to $n=199$ so that it reaches $E=30$ at all; without that enlargement it stops
at $E\simeq12$.}
\label{tab:tabulation}
\begin{tabular}{rllll}
\toprule
 & \multicolumn{2}{c}{binomial tables} & \multicolumn{2}{c}{logarithms
 (\code{tmb\_kam})}\\
\cmidrule(lr){2-3}\cmidrule(lr){4-5}
$E$ & $|\text{lad}-\text{cls}|$ & $\varepsilon$
    & $|\text{lad}-\text{cls}|$ & $\varepsilon$\\
\midrule
12 & $9.9\times10^{-12}$ & $2.4\times10^{-11}$ & $4.2\times10^{-14}$ & $8.3\times10^{-14}$\\
20 & $7.9\times10^{-7}$  & $1.5\times10^{-6}$  & $6.9\times10^{-13}$ & $1.6\times10^{-12}$\\
24 & $1.3\times10^{-4}$  & $2.7\times10^{-4}$  & $2.6\times10^{-12}$ & $5.9\times10^{-12}$\\
28 & $7.3\times10^{-2}$  & $1.4\times10^{-1}$  & $1.3\times10^{-11}$ & $3.8\times10^{-11}$\\
30 & $6.4\times10^{-1}$  & $5.0\times10^{-1}$  & $7.9\times10^{-11}$ & $8.9\times10^{-11}$\\
\bottomrule
\end{tabular}
\end{table}

\begin{table}[htb]
\centering
\caption{Three-particle CFPs obtained by diagonalizing $\hat\Lambda$ of
Eq.~(\ref{eq:lambda}) in its two $\hat P_{12}$ parity sectors, with the
required $d=\tfrac13$ brackets supplied by each of the three routes.
$\varepsilon$ is the departure of Eq.~(\ref{eq:interr}) of any eigenvalue from
the exact integers $\pm2,\pm1$. The cost of producing the brackets is not
repeated here --- it is Table~\ref{tab:speed} --- and the diagonalization
column is the same for all three routes, since they differ only in the matrix
handed to it. That column is an OpenBLAS \code{dsyev} timing on the two parity
sectors, measured for every row in a single uncontended run. Above $E=30$ we ran only the SU(3)-scheme routes: the VCS route
returns non-finite values for every block with $E+L\ge60$ and so fails
throughout, while the classical route becomes prohibitively expensive
(Sec.~\ref{sec:cost}). No block of the classical route returned a non-finite
value anywhere in the range tabulated.}
\label{tab:cfp}
\small
\begin{tabular}{rrr lll}
\toprule
 & & diag. & \multicolumn{3}{c}{$\varepsilon$}\\
\cmidrule(lr){4-6}
$E$ & $n_{\max}$ & [s] & ladder & VCS & Kam01\\
\midrule
 8 &  30 & 0.000 & $2.0\!\times\!10^{-14}$ & $2.5\!\times\!10^{-13}$ & $2.6\!\times\!10^{-14}$\\
12 &  75 & 0.001 & $6.9\!\times\!10^{-14}$ & $4.4\!\times\!10^{-11}$ & $8.3\!\times\!10^{-14}$\\
16 & 147 & 0.002 & $2.8\!\times\!10^{-13}$ & $9.1\!\times\!10^{-9}$  & $4.5\!\times\!10^{-13}$\\
20 & 252 & 0.006 & $1.4\!\times\!10^{-12}$ & $2.7\!\times\!10^{-6}$  & $1.6\!\times\!10^{-12}$\\
24 & 405 & 0.019 & $5.8\!\times\!10^{-12}$ & $9.2\!\times\!10^{-4}$  & $5.9\!\times\!10^{-12}$\\
28 & 605 & 0.057 & $6.6\!\times\!10^{-11}$ & $5.2\!\times\!10^{-2}$  & $3.8\!\times\!10^{-11}$\\
30 & 726 & 0.091 & $6.3\!\times\!10^{-10}$ & $4.5\!\times\!10^{-1}$  & $8.9\!\times\!10^{-11}$\\
\midrule
35 & 1092 & 0.295 & $9.5\!\times\!10^{-9}$  & fails & ---\\
40 & 1575 & 0.835 & $5.6\!\times\!10^{-8}$  & fails & ---\\
45 & 2176 & 2.233 & $1.4\!\times\!10^{-6}$  & fails & ---\\
50 & 2907 & 5.639 & $4.5\!\times\!10^{-6}$  & fails & ---\\
\bottomrule
\end{tabular}
\end{table}

The three routes separate into a clear speed--accuracy pattern, and it is not
the one a reading of Sec.~\ref{sec:cost} alone would suggest.

On \emph{accuracy} the classical route and the ladder route are close over the
whole range: the two $\varepsilon$ columns agree to within a factor of two
through $E=28$, and at $E=30$ the classical value is smaller by a factor of
seven ($8.9\times10^{-11}$ against $6.3\times10^{-10}$). Neither is limited by
its tabulation any longer, so what separates them is ordinary rounding, and
the classical route --- which shares nothing between elements --- accumulates
slightly less of it as the block dimension grows.

On \emph{cost} the margin is enormous. At $E=30$ the ladder route produces the
shell's brackets in $0.30$~s against $310$~s
(Table~\ref{tab:speed}, prepare plus evaluate in both cases): a factor of
about $1000$. That gap is what stops the classical route well short of $E=50$.
Its cost per block grows as $n^{2}$ times a per-element cost that itself grows
with $E$; the whole $E=32$ shell already takes ten minutes, and a single
bracket at $E=50$ takes $0.37$~ms, so the $975\times975$ block at $L=2$ alone
would need six minutes against $0.05$~s for the ladder. Extending the
tabulation to $E=50$ by that road is a matter of machine-days, which is
precisely the practical point: the classical formula remains the accuracy
reference, but it is not a tool for sweeping a model space. The accuracy the
classical route buys with that time is not required --- $6\times10^{-10}$
already identifies every eigenvalue with its exact integer beyond any
ambiguity, since the integers are separated by one. Carried to $E=50$ the
ladder route degrades to $4.5\times10^{-6}$ --- still five orders of margin
against the nearest competing integer --- while producing every bracket of
that shell in $13$~s and diagonalizing all $51$ of its blocks in a further
$5.1$~s.

The practical reading is therefore straightforward. The classical formula,
correctly tabulated, remains the definitive check on the brackets, and this is
the role it plays in Sec.~\ref{sec:test}; but it costs a factor of a thousand
and yields no better answer here. Where brackets are needed block by block
across a model space --- the situation in the applications these brackets
exist for --- the ladder construction is the route to use. What the table
rules out is the VCS library as a middle ground: it is slower than the ladder
\emph{and} several orders less accurate, and has lost the symmetry
classification entirely by $E=28$.

\section{Summary}
\label{sec:summary}
\code{SU3HOB-ladder} computes general Talmi--Moshinsky harmonic-oscillator
brackets in the $\su{SU}(3)$ basis by the method of Ref.~\cite{KALINAUSKAS2025123033},
obtaining the $\su{SU}(3)\supset\su{SO}(3)$ isofactors from the $\su{U}(2)$
pseudo-spin generators of the same group chain. Complementarity of the
$\su{U}(3)$ and $\su{U}(2)$ labels inside $[E]$ of $\su{U}(6)$ makes the
highest-weight states the null space of the raising operator, with the
remaining members of each multiplet obtained by lowering; the resulting
isofactors are orthonormal by construction, and no $\su{SU}(3)$
Clebsch--Gordan machinery, $K$-matrix or outer-multiplicity resolution appears
anywhere.

Measured against three independent routes to the same brackets --- the same
scheme with isofactors from the vector-coherent-state library
\Su{}~\cite{Bahri_2004}, and the classical Talmi--Moshinsky sum
\cite{Kamuntavicius,BuckMerchant} evaluated in logarithmic form --- the
ladder construction is the fastest of the three and the only one usable over
the whole range tested. The SU(3) scheme itself is worth two to three orders
of magnitude over element-wise evaluation of the classical formula; within it,
the ladder route is a further factor six to eight faster than the library
route and several orders of magnitude more accurate, retaining
$\max|\mathrm{HOB}\cdot\mathrm{HOB}^{T}-I|\lesssim2\times10^{-6}$ at $E=50$.
The comparison also isolates where each alternative stops: the classical
formula is limited by how its combinatorial factors are tabulated rather than
by the formula, and by its cost; the library route returns non-finite values
for every block with $E+L\ge60$.

One qualification belongs with that summary. Accuracy and cost do not order
the three routes the same way. Once its factorials are held in logarithmic
form, the classical formula matches the ladder construction digit for digit
over the whole range where both can be run --- the two agree to within a
factor of two at every shell of Table~\ref{tab:cfp} --- and it pays for that
with a factor of roughly $1000$ in time. It is the right instrument for
verifying brackets and the wrong one for producing them in quantity, which is
the division of labor adopted throughout this paper: the classical route
validates in Sec.~\ref{sec:test}, the ladder route computes.

For codes that already link \Su{} for symmetry-adapted bases, the interface
documented in Sec.~\ref{sec:protocol} --- the
\code{readfact}/\code{readtab}/\code{cgu3hw}/\code{cgu3o3} protocol, the single
state-dependent phase $(-1)^{(e_1-l_1)/2+(e_2-l_2)/2}$ relating the VCS and HO
conventions, and the analytic treatment of the scalar $(0,0)$ factor ---
remains available as an alternative isofactor source, and served here as the
independent check on the ladder construction.

\section*{Code availability}

\code{SU3HOB-ladder} v1.0 is released under the MIT license and is available at
\url{https://github.com/Augustinaz/SU3HOB-ladder}. The v1.0 release measured
here is archived at
\href{https://doi.org/10.5281/zenodo.22150398}{10.5281/zenodo.22150398}; the
concept DOI \href{https://doi.org/10.5281/zenodo.22150397}{10.5281/zenodo.22150397}
resolves to the latest version. The distribution contains
the library, the command-line front end, the validation driver with
its stored reference output, and the benchmark driver, which reproduces the
ladder columns of Tables~\ref{tab:speed} and~\ref{tab:accuracy} with
\code{make bench} on any machine. The derivation of the ladder construction is
Sec.~\ref{sec:raising} above and is not repeated in the distribution. The
measurement logs behind every table in this paper are included, so the figures
quoted here can be checked directly; reproducing the VCS columns in addition
requires \Su{}~\cite{Bahri_2004}, which is not redistributed here.

\section*{Acknowledgments}
We thank the authors of Ref.~\cite{Bahri_2004} for the \Su{} library, which
provided the independent isofactor route used here for validation.

\section*{Funding}

This research did not receive any specific grant from funding agencies
in the public, commercial, or not-for-profit sectors.

\section*{Declaration of competing interest}

The authors declare that they have no known competing financial interests
or personal relationships that could have appeared to influence the work
reported in this paper.


\end{document}